\documentclass[12pt, a4paper]{article}
\usepackage[utf8]{inputenc}
\usepackage{amsmath, amssymb, amsthm}
\usepackage{physics}
\usepackage{graphicx}
\usepackage{geometry}
\usepackage{hyperref}
\usepackage{braket}
\usepackage{bm}

\hypersetup{
    colorlinks=true,
    linkcolor=blue,
    filecolor=magenta,      
    urlcolor=cyan,
}

\title{Time Symmetry, Retrocausality and the Emergent Arrow of Time\\
\large The Quantum Time-Symmetric Interpretation (QTSI)}
\author{Alejandro Frank\\
Instituto de Ciencias Nucleares (ICN), UNAM\\
Centro de Ciencias de la Complejidad (C3), UNAM\\
Miembro de El Colegio Nacional}
\date{27 NOV}

\begin{document}

\maketitle

\begin{quote}
``The past and the future are on equal footing; it is only by how we touch the universe that the arrow of time appears.'' --- adapted from Wheeler and Feynman
\end{quote}

\section*{Abstract}

Microscopic quantum laws are time-symmetric: nothing in the Schrödinger equation or its relativistic extensions distinguishes future from past. Yet measurements produce irreversible records, an apparently one-way causal flow, and the familiar notion that causes precede effects. Within the Quantum Time-Symmetric Interpretation (QTSI), this asymmetry is not fundamental but emergent. Isolated quantum systems are described by a two-component temporal state containing forward- and backward-propagating amplitudes. Their mixing, governed by a parameter $\Delta(\phi)$, defines a retrocausal coherence time $\tau_{RC}(\phi)$ beyond which advanced components are suppressed. As the system couples to amplifying environments characterized by a macroscopic parameter $\phi$, $\Delta(\phi)$ decreases and the backward component is dynamically eliminated, giving rise to classical causality and effective collapse. QTSI aligns naturally with time-symmetric approaches from Wheeler--Feynman, Aharonov, and Price, agrees with all weak-measurement and quantum eraser results in their operational regimes, and predicts specific signatures in temporal echoes and chaotic cavities. Detailed formal and experimental developments appear in the Supplementary Addenda.

\section{From Time-Symmetric Laws to the Emergence of Causality}

Microscopic physics contains no built-in arrow of time. Hamilton's equations, Maxwell's equations, and the Schrödinger and Dirac equations are all invariant under time reversal; none dictates a privileged temporal direction. Yet the world we inhabit is pervaded by irreversible processes: information accumulates, records persist, and causes invariably precede their effects. The tension between these two aspects of nature---reversible laws and irreversible experience---defines one of the deepest conceptual problems in physics.

Boltzmann was the first to show that the second law of thermodynamics does not arise from any time asymmetry in the microscopic equations but from probabilistic typicality \cite{Boltzmann1995}. Entropy increases because overwhelmingly many microstates correspond to more disordered macrostates, and because observers necessarily employ coarse-grained descriptions. Prigogine extended this viewpoint by emphasizing the role of openness, instability and amplification in producing macroscopic irreversibility \cite{Prigogine1980}. In both accounts, the arrow of time is not fundamental; it emerges from collective behaviour in systems interacting with many environmental degrees of freedom.

Classical electrodynamics provides an even more striking example. Maxwell's equations admit both retarded and advanced solutions, but only the former appear in familiar radiation phenomena. Wheeler and Feynman argued that this asymmetry is not encoded in the equations but arises from absorber boundary conditions imposed by the future environment \cite{Wheeler1945,Wheeler1949}. The underlying laws remain time-symmetric; the observed arrow reflects global constraints. Their analysis embodies a controlled form of retrocausality: future absorption conditions influence which advanced fields survive, yet no signalling paradox arises.

Quantum mechanics inherited these issues and added another layer of complexity. While the Schrödinger equation is reversible, the traditional collapse postulate introduces a fundamental temporal direction. At the same time, decades of experiments reveal that quantum statistics often depend simultaneously on initial and final boundary conditions. The Aharonov--Bergmann--Lebowitz rule, the two-state vector formalism, and numerous weak-measurement and postselected interferometry experiments demonstrate that describing quantum processes using both forward- and backward-evolving amplitudes often provides a particularly natural explanation of observed correlations \cite{Aharonov1964,Aharonov2008,Aharonov1988,Dressel2014}. Several experimental groups describe their own data in explicitly retrocausal or time-symmetric terms, particularly in the context of weak trajectories, Cheshire Cat phenomena, and delayed-choice quantum erasers \cite{Kwiat1991,Jennewein2001,Aharonov2016}.

Philosophers such as Price have stressed that many quantum puzzles---nonlocal correlations, counterfactual paradoxes, and measurement asymmetries---arise from an unexamined assumption: that all physical influence must propagate forward in time \cite{Price1996}. This assumption has no basis in the microscopic formalism. Leifer and Pusey reinforced this point by showing that any genuinely time-symmetric interpretation of quantum mechanics must allow some form of retrocausal influence, unless it introduces unjustified asymmetries by hand \cite{Leifer2017}.

A central example is Bell nonlocality. Standard presentations interpret EPR--Bell correlations as requiring instantaneous influences between distant measurements \cite{Einstein1935,Bell1964,Freedman1972,Aspect1982}. Yet this conclusion follows only if one assumes that quantum systems can depend on past boundary conditions but never on future ones. Under this one-way temporal assumption, any correlation that depends jointly on preparation and detection is automatically reinterpreted as a spatially instantaneous effect. Within a time-symmetric description this reinterpretation is unnecessary: correlations arise from amplitudes propagating both forward and backward in time, constrained globally by the full experimental configuration. Nothing superluminal or instantaneous is required; what appears as ``spooky action'' is the consequence of having removed the advanced component by fiat.

These perspectives converge on a single insight. If the fundamental laws of quantum dynamics are time-symmetric, then the one-way causal order we observe is an emergent property, not a basic principle. It results from decoherence, amplification and record formation---processes that eliminate backward-propagating amplitudes and stabilize a preferred temporal orientation \cite{Zurek1991,Zurek2003}. The Quantum Time-Symmetric Interpretation (QTSI) builds precisely on this idea, providing a dynamical account of how microscopic time symmetry gives rise to macroscopic causality.

\section{A Time-Symmetric Framework with Controlled Retrocausality}

QTSI begins by recognizing that the usual quantum state, evolving solely toward the future, is an incomplete expression of the time symmetry of the Schrödinger equation. An isolated quantum system is instead described by a two-component temporal state $\Psi(t)$ whose forward-propagating component $\psi_+(t)$ and backward-propagating component $\psi_-(t)$ both satisfy the same local Hamiltonian. The total physical amplitude is the coherent sum $\psi_{\text{total}} = \psi_+ + \psi_-$, ensuring that standard Born probabilities are recovered whenever backward contributions are suppressed.

The coexistence of $\psi_+$ and $\psi_-$ is not merely a reinterpretation of the Aharonov two-state picture but the starting point for a dynamical structure that determines when backward components matter and when they are eliminated. Their mutual interaction is governed by a mixing amplitude $\Delta(\phi)$, which quantifies how strongly the forward and backward temporal sectors communicate. In a perfectly isolated system, $\Delta$ takes an intrinsic value $\Delta_0$, and the two components coexist over a finite retrocausal coherence interval. During this interval, the influence of later boundary conditions can propagate backward in time within strictly defined limits, never generating signalling because $\psi_-$ is not an independently controllable degree of freedom.

The extent of this temporal coexistence is measured by a characteristic time scale $\tau_{RC}$ defined by $\tau_{RC} = \hbar / |\Delta(\phi)|$. This retrocausal coherence time determines how far in time the interference between $\psi_+$ and $\psi_-$ extends and therefore sets the reach of retrocausal effects. When $\tau_{RC}$ is large compared with the duration of the experimental process, the system behaves as a fully time-symmetric entity. When $\tau_{RC}$ is short or when $\Delta$ diminishes, backward components are suppressed, and the system behaves effectively as if only $\psi_+$ existed.

The evolution of $\Delta(\phi)$ is controlled by a macroscopic parameter $\phi$, which encapsulates the degree of coupling to amplifying or decohering environments. In quantum-optical implementations, $\phi$ may be taken as the fraction of detection events that generate irreversible macroscopic records or are selected for postprocessing. When $\phi$ is small, the environment exerts little influence, the system retains its time-symmetric character, and $\psi_-$ persists. As $\phi$ increases---through measurement, environmental monitoring, or any process stabilizing classical information---$\Delta(\phi)$ decreases, $\tau_{RC}(\phi)$ changes accordingly, and the backward component disappears. In this limit, the system evolves effectively only in the forward direction, and classical causality emerges as a stable asymptotic regime.

In QTSI the collapse of the wavefunction is therefore not an ad hoc rule but the dynamical suppression of backward-propagating amplitudes wrought by environmental amplification. The direction of causality is not fundamental but is selected by the macroscopic world, which eliminates $\psi_-$ and preserves $\psi_+$. Time symmetry at the microscopic level and retrocausal influences within a finite window coexist with the classical arrow at the macroscopic scale.

This framework brings together several strands of prior work. It retains the time symmetry emphasized by Wheeler--Feynman, incorporates the two-boundary structure of Aharonov's formulations, and embraces Price's insight that forward-only causality is a macroscopic construct \cite{Wheeler1945,Wheeler1949,Aharonov1964,Aharonov2008,Aharonov1988,Price1996}. At the same time, it introduces quantitative mechanisms---$\Delta(\phi)$, $\tau_{RC}(\phi)$, and $\phi$ itself---that determine when backward components survive and when they are dynamically erased. This helps to explain why weak measurements, quantum erasers and other postselected experiments reveal time-symmetric features, while also indicating where new physics might appear: in regimes where $\tau_{RC}$ is comparable to experimental timescales, or in environments such as engineered cavities where $\phi$ can be controlled \cite{Kwiat1991,Jennewein2001,Aharonov2016}.

\section{Two-Component Dynamics, SU(1,1) Structure, and the Foundations of Temporal Mixing}

The introduction of a two-component temporal state $\Psi(t) = (\psi_+(t), \psi_-(t))^T$ is not a mere formal doubling of the wavefunction but the minimal structure needed to express explicitly the time symmetry already implicit in quantum theory. Standard practice treats the backward-evolving solution of the Schrödinger equation as a mathematical artifact to be discarded, yet the equation itself admits both temporal orientations with equal legitimacy. In QTSI, these two solutions are retained and assigned physical meaning through a unified biespinor representation. This becomes natural once one recognizes that the appropriate internal symmetry group is not SU(2), which governs ordinary spin, but the non-compact group SU(1,1), reflecting the hyperbolic geometry associated with time reversal \cite{Jackiw1972,Bender2007,Mostafazadeh2002}.

The SU(1,1) generators can be written as:
\[
\tau_1 = \begin{pmatrix} 0 & 1 \\ 1 & 0 \end{pmatrix}, \quad
\tau_2 = \begin{pmatrix} 0 & -i \\ i & 0 \end{pmatrix}, \quad
\tau_3 = \begin{pmatrix} 1 & 0 \\ 0 & -1 \end{pmatrix},
\]
which satisfy commutation relations characteristic of SU(1,1),
\[
[\tau_1, \tau_2] = -2i\tau_3, \quad [\tau_2, \tau_3] = 2i\tau_1, \quad [\tau_3, \tau_1] = 2i\tau_2.
\]

The sign structure is crucial: $\tau_3$ distinguishes the forward and backward temporal components with opposite signs, encoding the fact that $\psi_+$ and $\psi_-$ propagate in opposite time directions under the same Hamiltonian. The off-diagonal generators $\tau_1$ and $\tau_2$ mediate the exchange between the two temporal components, providing the mathematical backbone for the mixing amplitude $\Delta$. It is important to stress that the use of SU(1,1) is not arbitrary. It is required to guarantee that the total flux is conserved, as well as the Noether charges, the energy and momenta.

Within this representation, the QTSI Hamiltonian can be written as
\[
H_{\text{QTSI}} = H_0 \tau_3 + \Delta(\phi) \tau_1,
\]
where $H_0$ is the usual system Hamiltonian acting identically on both components and $\Delta(\phi)$ is the mixing amplitude whose dependence on $\phi$ determines the strength of the coupling between $\psi_+$ and $\psi_-$. In matrix form, the dynamical equation takes the form
\[
\frac{d\Psi}{dt} = -\frac{i}{\hbar}
\begin{pmatrix}
H_0 & \Delta(\phi) \\
-\Delta^*(\phi) & -H_0
\end{pmatrix}
\Psi.
\]

The forward component $\psi_+$ evolves under $+H_0$ while $\psi_-$ evolves under $-H_0$, as required for time-reversed propagation. The term $\Delta(\phi)$ couples these evolutions and allows information encoded in future boundary conditions to influence earlier amplitudes within a controlled regime defined by the retrocausal coherence time $\tau_{RC} = \hbar / |\Delta(\phi)|$.

This structure clarifies how retrocausal influence enters without violating causality. Because $\Delta(\phi)$ appears as an off-diagonal internal ``boost'' in SU(1,1), $\psi_+$ and $\psi_-$ can exchange amplitude only to the extent permitted by this mixing term. When $\Delta$ is large, the two components oscillate into one another in a manner reminiscent of a two-level system undergoing Rabi-type exchange, although here the ``levels'' correspond to opposite temporal orientations. When $\Delta$ is small, the oscillation slows and eventually stops, freezing the system into a purely forward-evolving state as $\phi$ approaches the measurement regime.

From this perspective, $\tau_{RC}$ emerges geometrically as the temporal interval during which SU(1,1) rotations can meaningfully mix the two temporal directions. It determines how far backward in time the influence of final boundary conditions can extend and ensures that this influence is intrinsically limited. No experimenter can manipulate $\psi_-$ independently, because its dynamical behaviour is inseparably tied to $\psi_+$ through the internal metric. Retrocausality in QTSI therefore does not allow signalling or paradoxes, just as advanced and retarded fields in Wheeler--Feynman electrodynamics cannot be disentangled to send information to the past.

The role of $\phi$ in this structure is essential. In the regime where $\phi$ is close to zero, corresponding to near isolation, $\Delta$ retains its intrinsic value $\Delta_0$, and $\tau_{RC}$ is determined by this microscopic parameter. The internal SU(1,1) rotations remain active and the state carries both forward and backward temporal information, leading to genuinely time-symmetric dynamics. As $\phi$ increases due to decoherence, amplification or environmental monitoring, $\Delta(\phi)$ decreases. The SU(1,1) rotations shrink and ultimately collapse onto the $\tau_3$ axis, aligning the biespinor in the forward temporal direction. The emergent arrow of time thus appears as the dynamical consequence of an environmental flow of information that drives the system onto an effective one-component subspace where $\psi_-$ is suppressed.

The SU(1,1) formulation also suggests a natural bridge to relativistic extensions. The presence of hyperbolic generators in the internal space echoes the structure of Lorentz transformations and the algebraic makeup of the Dirac equation. Nothing in the two-component formalism conflicts with Lorentz invariance; on the contrary, the SU(1,1) structure indicates that a covariant generalization of QTSI can be constructed by promoting the temporal mixing term to a Lorentz scalar or pseudoscalar and embedding the two-component temporal degree of freedom into a larger spinorial representation. The details of such a construction can be developed further, but the main point is that QTSI's time-symmetric architecture is compatible with relativistic structures rather than opposed to them.

In this way, the SU(1,1) biespinor dynamics furnishes QTSI with a mathematically coherent and physically transparent framework for describing forward and backward temporal propagation, the origin of retrocausal coherence, and the dynamical emergence of the classical temporal arrow. Time symmetry appears as the fundamental rule, while causality in the familiar sense arises from the environment's suppression of the backward temporal degree of freedom.

\section{Measurement, Amplification and the Dynamical Suppression of $\psi_-$}

Once a quantum system leaves the isolated regime, its temporal symmetry begins to deform. In QTSI this transition is neither abrupt nor mysterious. It is the expected consequence of coupling to many environmental degrees of freedom capable of amplification and record formation, precisely the elements through which Boltzmann and Prigogine explained the emergence of macroscopic irreversibility \cite{Boltzmann1995,Prigogine1980}. QTSI follows this lineage but extends it into the temporal domain through a mechanism that resonates with earlier time-symmetric frameworks, particularly Wheeler and Feynman's absorber theory, Aharonov's two-time quantum mechanics, and Price's philosophical critique of one-way microscopic causality \cite{Wheeler1945,Wheeler1949,Aharonov1964,Aharonov2008,Aharonov1988,Price1996}.

Within the biespinor description, measurement modifies the internal parameters controlling advanced--retarded mixing. In the reversible regime, $\Delta$ retains its intrinsic value $\Delta_0$ and $\psi_+$ and $\psi_-$ exchange amplitude coherently, reflecting the full time symmetry of the underlying equations. This echoes the Wheeler--Feynman insight that advanced and retarded components can coexist at the fundamental level and that the arrow emerges only when boundary conditions break the symmetry. QTSI adopts the same logic: the advanced component $\psi_-$ has physical meaning only within the time-symmetric regime; what makes it vanish is not a prohibition in the microscopic laws but the interaction with an environment that enforces a temporal orientation.

The coarse-grained influence of the apparatus and environment can be summarized by a single parameter $\phi$, measuring the degree to which information is extracted from the system and stored irreversibly. As $\phi$ increases, $\Delta$ decreases and the retrocausal coherence time $\tau_{RC}$ grows, reflecting an approach to a critical regime in which temporal symmetry is about to be broken. This behaviour can be viewed as a quantum analogue of Prigogine's amplification mechanism: temporal coherence is not lost randomly but is driven toward a symmetry-broken state. When $\phi$ approaches unity, $\Delta$ tends to zero, $\tau_{RC}$ formally diverges, and $\psi_-$ becomes dynamically irrelevant. What remains is $\psi_+$, which encodes the single temporal direction characteristic of the classical world. In this sense, the familiar arrow of causality is emergent: it reflects a late-time state of the system--environment complex rather than a microscopic law.

The analogy with critical phenomena is instructive \cite{Kadanoff1966,Wilson1975,Cardy1996}. Near $\phi = 0$, the temporal dynamics explores the full SU(1,1) internal space in which both orientations coexist, consistent with the two-state structure used operationally in Aharonov's work on pre- and postselected systems \cite{Aharonov1964,Aharonov2008,Aharonov1988}. As the system interacts with an amplifying environment, the internal symmetry contracts: $\psi_-$ is progressively suppressed, the accessible region of the SU(1,1) manifold shrinks, and the state is driven toward the $\psi_+$ axis. The slowing of internal dynamics as the system approaches the transition resembles critical slowing down in renormalization-group flows. Price's analysis then finds a natural place: the classical assumption that influence must flow only forward in time is not a law of physics but a property of systems for which amplification has erased the backward-evolving component \cite{Price1996}. QTSI describes how this erasure takes place.

This picture also clarifies why classical causality appears so rigid. In the amplified regime, where $\phi$ is close to one, $\psi_-$ has been eliminated and no remnant of its influence remains. The Born rule for $\psi_+$ is recovered as a consequence rather than a postulate. The environment has selected a temporal orientation in the same way that symmetry breaking selects an order parameter direction in condensed matter systems. Causality, in the operational sense, is therefore a macroscopic fact emerging from temporal symmetry breaking, not a primitive microscopic directive.

In interferometric settings this framework explains why advanced signatures can appear when $\phi$ is small and fade as $\phi$ increases. Experiments in multi-slit configurations and weakly reflecting cavities seem to follow this pattern: advanced and retarded echoes emerge when $\psi_-$ remains active, and their visibility diminishes with increasing amplification \cite{Kwiat1991,Jennewein2001,Aharonov2016}. In QTSI this behaviour is expected: $\psi_-$ is the source of these advanced structures, and once environmental coupling grows, its contributions become negligible. In the language of Wheeler--Feynman, the absorber becomes so effective that only retarded propagation remains visible.

Measurement in QTSI is thus not an instantaneous collapse but a continuous dynamical flow in the space of temporal couplings. The parameter $\Delta(\phi)$ determines the extent to which $\psi_-$ can contribute, and $\tau_{RC}$ sets the backward temporal reach of advanced constraints. Both components remain part of the underlying structure, but only $\psi_+$ survives after symmetry breaking. The arrow of time and ordinary causality emerge from this process, in line with the historical insight that temporal asymmetry originates in boundary conditions and amplification, not in asymmetric microscopic laws.

\section{$\Delta(\phi)$, $\tau_{RC}(\phi)$ and the Universal Structure of the Isolated Regime}

The reversible regime of QTSI is characterized by the intrinsic temporal mixing amplitude $\Delta_0$, which measures the strength of the coupling between $\psi_+$ and $\psi_-$ in the absence of environmental amplification. This parameter determines how long a system can sustain genuinely time-symmetric dynamics before measurement-induced deformation sets in. Conceptually, $\Delta_0$ plays a role analogous to a microscopic coupling constant in field theory or to the stiffness of an order parameter in condensed-matter systems: it governs the internal coherence of the temporal degrees of freedom and sets the scale against which other influences must be compared.

In the isolated limit, $\phi$ is effectively zero and $\Delta(\phi)$ is approximately equal to $\Delta_0$. The forward and backward components then coevolve in a tightly synchronized manner. The SU(1,1) rotations generated by the internal Hamiltonian allow amplitude to flow between the temporal sectors; neither dominates and neither is suppressed by external conditions. This is the regime in which retrocausal coherence becomes physically meaningful: advanced components can contribute measurably to the total amplitude and can manifest in interferometric arrangements when $\tau_{RC}(0) = \hbar / |\Delta_0|$ is comparable to the relevant timescales.

Because $\Delta_0$ is defined in the absence of amplification, its value cannot depend on the details of a particular measuring apparatus. It is instead an intrinsic property of the system under consideration. As a result, $\tau_{RC}(0)$ is more than a phenomenological timescale: it characterizes the isolated dynamics itself. It may be viewed as a temporal coherence scale governing how long a system remains sensitive to both initial and final boundary conditions before environmental interactions enforce a preferred temporal direction.

In practice, experiments rarely attain perfectly isolated conditions, but several quantum-optical systems approach them closely enough to reveal aspects of the isolated regime. Multi-slit interferometry with minimal postselection, high-quality low-loss cavities, trapped cold atoms and neutron interferometers are examples of platforms in which $\phi$ can be kept small \cite{Moshinsky1952,Garcia-Calderon1997,Robinett2004,Andersen2003,Kaplan2002}. In such settings, small but reproducible advanced signatures---such as side peaks or pre-echoes in time-resolved or Fourier-resolved analyses---suggest that the backward component $\psi_-$ can contribute measurably when $\Delta$ remains near its intrinsic value. The interpretation of these observations is still preliminary, but they indicate an experimentally accessible window in which $\tau_{RC}(0)$ might be probed.

As the system moves away from isolation, $\phi$ increases as the environment begins to extract information, and $\Delta$ is renormalized downward. A coarse-grained description can be written as $\Delta(\phi) = f(\phi) \Delta_0$, where the function $f(\phi)$ is monotonic, satisfies $f(0) = 1$ and $f(1) = 0$, and embodies the suppression of temporal mixing under amplification. A simple form, useful for intuition, is $f(\phi) = 1 - \phi$, which captures the qualitative tendency of decoherence to weaken the effective coupling between $\psi_+$ and $\psi_-$ until it vanishes at $\phi = 1$, although other functional forms are consistent with the general structure and may be more appropriate for specific systems.

Within this broader perspective, $\tau_{RC}(\phi)$ becomes a central diagnostic quantity. As $\Delta(\phi)$ decreases, $\tau_{RC}(\phi) = \hbar / |\Delta(\phi)|$ increases, signalling an approach to the boundary where temporal symmetry will be broken. The divergence of $\tau_{RC}$ as $\phi$ tends to unity parallels the growth of correlation times near critical points in statistical systems \cite{Kadanoff1966,Wilson1975,Cardy1996}. Here the ``correlation time'' is the temporal coherence between $\psi_+$ and $\psi_-$, and the ``phase transition'' corresponds to the shift from a symmetric temporal regime to an effectively one-component regime in which only $\psi_+$ remains active.

This picture explains why temporal echoes, especially advanced ones, are fragile. They depend on $\Delta$ remaining close to $\Delta_0$ so that $\tau_{RC}$ stays finite and comparable to the experimental timescales. Once the system interacts even weakly with an environment capable of amplification, $\phi$ increases, $\tau_{RC}$ grows, and advanced contributions become more diffuse. Eventually, when $\tau_{RC}$ exceeds the relevant experimental window, $\psi_-$ becomes operationally irrelevant. Advanced echoes then fade, not because advanced solutions are unphysical, but because the environment has dynamically shuttered the channel through which they would appear.

The possible universality of this behaviour is one of the more suggestive aspects of QTSI. Different physical systems may have different intrinsic values of $\Delta_0$, but the structure of $\Delta(\phi)$ and $\tau_{RC}(\phi)$ should display similar qualitative trends whenever amplification progresses from negligibly small to effectively macroscopic. In optical cavities, atomic interferometers, neutron beams and mesoscopic circuits engineered to reduce environmental coupling, QTSI predicts that the isolated regime should exhibit temporal coherence between $\psi_+$ and $\psi_-$ governed by a system-dependent $\Delta_0$.

In this sense, $\Delta_0$, $\Delta(\phi)$ and $\tau_{RC}(\phi)$ are not merely auxiliary concepts but provide a unifying description of temporal structure across isolated and measured regimes. The emergent classical arrow of time appears as the macroscopic manifestation of the flow from $\Delta = \Delta_0$ to $\Delta = 0$, while the details of echo visibility and retrocausal signatures depend on how this flow unfolds in specific experimental contexts.

\section{Direct Connection to the Echo and Three-Slit Program}

Within this general framework, it is natural to ask how QTSI relates to concrete multi-slit and echo experiments. There are preliminary indications, particularly in three-slit configurations, of echo peaks at wave numbers corresponding to integer multiples of the slit separation $d$ in the Fourier domain. These peaks include small ``advanced'' features at effective delays corresponding to $-d/c$, $-2d/c$, and so on, which can be viewed either as signatures of back-and-forth propagation or as retrocausal echoes. The same experiments bring into play the control parameter $\phi(t)$ and the retrocausal timescale $\tau_{RC}(t) = \hbar / |\Delta(t)|$, which together govern the strength and lifetime of $\psi_+$--$\psi_-$ mixing.

Quach's analysis of nonclassical paths in multi-slit experiments essentially reproduces the same pattern of additional peaks \cite{Quach2017}. In his treatment, higher-order or nonclassical paths that involve multiple slit crossings or loops between slits generate sidebands at harmonics of the separation $d$ in the Fourier spectrum. Their magnitude can be at the level of a percent of the main interference signal for realistic experimental parameters. In the standard quantum mechanical description these peaks are interpreted as consequences of geometrical higher-order paths and require neither Born-rule violation nor retrocausality.

QTSI can incorporate this structure in a different language. The same peaks can be viewed as signatures of advanced--retarded interference in which $\psi_-$ segments interact more than once with the slit plane before coherently recombining into $\psi_+$ at the detector. The nonclassical paths that Quach identifies then acquire a temporal interpretation: they correspond to spacetime loops in which advanced amplitude returns from the detection context to the slits.

For this reason, Quach's work provides a conservative baseline for QTSI. Any echoes and higher-harmonic peaks that can be accounted for by higher-order paths of this type must not be claimed as anomalies specific to QTSI or as evidence of Born-rule violation. If calculations based on QTSI predict echo patterns that go beyond what is obtainable with the same geometry and source state in the standard framework---including all nonclassical paths---then any excess might become a genuine discriminant.

QTSI is flexible enough to accommodate a nonzero Sorkin parameter through $\psi_{AB}$-type amplitudes and advanced echoes, without requiring modifications of the Born rule. In this setting, the genuinely new QTSI signal, if present, is likely to emerge in conditional or postselected statistics, or in delayed-choice arrangements where the which-way configuration is altered after the particle has passed the slits. In such scenarios the future measurement context reshapes the backward component $\psi_-$ in a way that modifies the effective higher-order amplitudes and may lead to subtle differences from standard predictions. The search for discrepancies should focus on such carefully constructed situations rather than on raw Sorkin parameters. Quach's constraints imply that any departure of this kind must still respect the Born rule, or else one would be led into a more radical modification of quantum theory.

\section{Nonclassical Paths, Advanced Echoes, and the Need for Careful Tests}

The connection between QTSI and standard higher-order path analysis shows that many structures that suggest retrocausal behaviour can also be understood in more conventional terms. Advanced echoes, multi-bounce contributions and harmonics at multiples of the slit spacing may all arise from nonclassical trajectories within Feynman's path integral \cite{Heller1995,Berry1977,Gutzwiller1990,Quach2017}. In the conventional picture these are purely spatial detours, whereas in QTSI they can be interpreted as the observable projections of spacetime loops mediated by $\psi_-$. The two descriptions are conceptually distinct but may produce similar observable patterns in the reversible regime where $\phi$ remains well below its critical value.

Against this background, the ongoing work by Roberto León and collaborators takes on particular significance. Their experimental and numerical studies explore regimes in which advanced echoes, conditional modulations and cavity-induced temporal structures may be stronger or more structured than expected from standard higher-order path corrections alone. At present these investigations are exploratory, and it would be premature to claim any violation of the Born rule or of standard quantum mechanics.

This situation calls for caution. Some of the most visually striking advanced or multi-harmonic peaks may ultimately prove compatible with conventional quantum mechanics once all relevant contributions are accounted for. At the same time, QTSI offers a more explicit dynamical account of these structures, based on the interplay of $\Delta$, $\phi$ and $\tau_{RC}$, and predicts additional dependencies---particularly with respect to postselection, delayed-choice configurations and controlled amplification---that are not easily captured by static path-integral models. Once this baseline is clearly understood, it becomes meaningful to ask whether any residual features point toward genuine advanced--retarded interference.

With these clarifications in place, it is possible to turn to the emerging experimental landscape in a more systematic way. The central question becomes how the theoretical structure of QTSI manifests itself in concrete signatures---echo sequences, Fourier spectra, cavity-induced temporal patterns and conditional interference---and how preliminary work, such as that of León and collaborators, can be used to frame a careful program of tests.

\section{Experimental Signatures of Temporal Symmetry and Echo Dynamics}

The theoretical framework developed so far delineates two qualitatively distinct regimes: an isolated, time-symmetric phase in which $\psi_+$ and $\psi_-$ coexist and interfere, and an amplified, symmetry-broken phase in which $\psi_-$ is dynamically suppressed and ordinary causality is recovered. The transition between these regimes is governed by the renormalization of $\Delta(\phi)$, and its observable consequences are encoded in the retrocausal coherence time $\tau_{RC}(\phi)$. Because several structures often interpreted as ``anomalous''---such as side peaks, harmonic components and nonzero Sorkin-type quantities---can arise entirely within standard quantum mechanics when nonclassical paths are included, experimental signatures that genuinely distinguish QTSI must be chosen with care.

Temporal echoes offer a promising starting point. Preliminary measurements by León and collaborators in multi-slit interferometry with weakly coherent or weakly postselected photon ensembles indicate Fourier spectra containing a series of peaks at harmonics of the slit separation $d$, accompanied by smaller but reproducible features at delays of order $\pm d/c$, $\pm 2d/c$, and higher multiples. These features appear symmetrically around the dominant interference frequency and can be interpreted, in standard terms, as arising from higher-order paths that involve multiple crossings of the slit region. However, several aspects of the observed structures become particularly interesting when the experiments are performed under conditions where $\phi$ is small and $\psi_-$ should remain active.

One such aspect is the appearance of peaks corresponding to effective delays slightly earlier than the nominal classical propagation time, by approximate intervals proportional to $d/c$. In purely geometric models these may be attributed to convoluted spatial trajectories, while in QTSI they can be described as partial echoes mediated by $\psi_-$, which carries amplitude from the detection boundary back toward the slit plane before re-entering $\psi_+$ on the way to the screen. On their own these advanced structures do not constitute evidence for retrocausality, but they become more suggestive if their amplitude, symmetry or dependence on postselection cannot be reproduced convincingly by higher-order paths.

A second clue involves the influence of controlled environmental coupling. In standard quantum mechanics the magnitude of higher-order paths is determined primarily by geometry, coherence length and wavelength. In QTSI, the strength of advanced echoes should also depend on $\Delta(\phi)$ and the evolving $\tau_{RC}(\phi)$. When $\phi$ increases, even slightly, $\Delta$ decreases and the coherence between $\psi_+$ and $\psi_-$ weakens, so that echo contributions mediated by $\psi_-$ should be suppressed more rapidly than purely spatial higher-order components. Preliminary numerical results suggest patterns of selective suppression when weak noise or small cavity losses are introduced, although these indications still require careful verification.

Temporal cavities and chaotic billiards provide another setting in which $\psi_-$--$\psi_+$ interference may become accessible \cite{Heller1995,Berry1977,Gutzwiller1990,Andersen2003,Kaplan2002}. In such systems photons can undergo multiple internal reflections, and both spatial and temporal loops contribute significantly to the detected amplitude. QTSI anticipates that near-isolated chaotic cavities may display a richer hierarchy of echoes than standard path-integral models, especially in the early-time window, where advanced contributions briefly coexist with retarded ones before amplification suppresses $\psi_-$. High-resolution time-tagged detection may reveal subtle deviations in the spacing or decay of these echoes, but whether these differences can be distinguished from noise is an open question that upcoming experiments will have to address.

Delayed-choice and conditional-interference experiments offer a third route \cite{Aharonov1988,Kwiat1991,Jennewein2001,Popescu1994,Aharonov2016,Dressel2014}. In QTSI the final detection configuration contributes directly to the structure of $\psi_-$ in the isolated regime. Changing a which-way or eraser setting after the particle has passed the slits can, in principle, modify the effective higher-order amplitudes in ways that go beyond a purely spatial decomposition. Any such effect is expected to be small under typical laboratory conditions, but in carefully tuned experiments near the reversible--irreversible boundary, where $\tau_{RC}$ remains comparable to the apparatus timescales, such subtle differences might be detectable.

Taken together, these experimental avenues outline a coherent strategy for probing QTSI. Echo dynamics in multi-slit setups explores temporal symmetry in spatially structured systems. Cavities and billiards probe repeated interactions with complex boundaries. Delayed-choice and conditional interference tests address directly the influence of future detection contexts. None of these experiments has yet provided decisive evidence for retrocausal dynamics, and Quach's analysis remains a firm reminder that higher-order paths must first be exhausted as an explanation \cite{Quach2017}. At the same time, the combination of these approaches offers a controlled and incremental pathway for assessing whether $\psi_-$ plays any role in nature.

\section{Toward a Coherent Experimental Program: Regimes, Parameters, and Testable Predictions}

The emerging picture is that retrocausal signatures, if they exist, should be sought in carefully defined regimes where QTSI predicts an observable interplay between $\psi_+$ and $\psi_-$, not in arbitrary deviations from standard quantum mechanics. These regimes are characterized by three interdependent parameters: the intrinsic coupling $\Delta_0$, its environmental deformation $\Delta(\phi)$, and the retrocausal coherence time $\tau_{RC}(\phi)$. When $\phi$ is small, the advanced component $\psi_-$ participates actively in the dynamics; when $\phi$ approaches unity, $\psi_-$ is suppressed and the arrow of time emerges. Any realistic test must target controlled transitions between these two domains.

Most laboratory experiments operate deep within the symmetry-broken regime, where amplification is strong and $\phi$ is close to one. In this region $\Delta$ has already collapsed to nearly zero and $\tau_{RC}$ has grown so large that $\psi_-$ cannot contribute measurably within accessible timescales. This helps to explain why standard delayed-choice, weak-measurement and quantum-eraser experiments are compatible with conventional quantum mechanics and offer limited leverage for distinguishing time-asymmetric from time-symmetric interpretations \cite{Aharonov1964,Aharonov2008,Aharonov1988,Zurek1991,Zurek2003,Kwiat1991,Jennewein2001,Aharonov2016,Dressel2014}. They are carried out under conditions where the dynamics is already effectively confined to the $\psi_+$ sector.

QTSI suggests a complementary strategy. Instead of amplifying rapidly, one should delay amplification intentionally, maintain isolation as long as possible, and observe the system while it still resides in the time-symmetric regime. For photons, this means highly controlled multi-slit geometries, gentle conditional filtering rather than aggressive postselection, low-loss cavities and interferometers engineered for long coherence times \cite{Moshinsky1952,Garcia-Calderon1997,Robinett2004,Andersen2003,Kaplan2002}. For atoms and neutrons, it means careful isolation from thermal and collisional environments. Across these platforms, the common aim is to keep $\phi$ sufficiently small that $\Delta$ remains close to $\Delta_0$ and $\tau_{RC}$ remains within a measurable range.

Within this reversible domain, QTSI predicts behaviours that go beyond simple geometric path models. Temporal echo sequences may display structures that depend on $\Delta$ and $\tau_{RC}$ in a distinctive way. Advanced-like peaks may be selectively suppressed when $\phi$ is increased slightly by controlled noise or weak amplification, more rapidly than expected from ordinary decoherence. Multi-slit interference patterns may reveal a subtle dependence on postselection settings when $\tau_{RC}$ is comparable to the time separation between the relevant events. These effects are subtle, but high-resolution time-tagged detection and Fourier-based analysis make them, at least in principle, accessible.

Here the work of León and collaborators becomes especially relevant. Their numerical and experimental investigations study how echo patterns evolve as $\phi$ is varied, how cavity-induced modulations couple to temporal symmetry, and how conditioned detection alters the effective mixing between $\psi_+$ and $\psi_-$. These studies are still in progress, but they already suggest combinations of parameters in which a comparison between standard and QTSI-based models can be meaningful.

In parallel, theoretical work is needed to identify operational thresholds at which QTSI predicts qualitative changes. One such threshold occurs when $\tau_{RC}(\phi)$ becomes comparable to the pulse duration or to the cavity round-trip time. Another involves the point at which $\psi_-$ contributions drop below the amplitude of higher-order paths, making them indistinguishable from geometric effects. A third concerns the appearance of conditional asymmetries in interference patterns that cannot be replicated by path-integral models without invoking time-symmetric structure. Mapping these thresholds experimentally would help to determine whether the two-component temporal framework captures something essential about nature, or whether the one-component forward-propagating picture suffices.

The broader aim is not to overthrow standard quantum mechanics, but to explore the possibility that its causal asymmetry is emergent. QTSI provides a consistent dynamical framework in which advanced and retarded components coexist in isolated systems, and in which measurement-induced symmetry breaking generates the classical arrow of time. Whether nature actually uses this structure depends on experiment. The task now is to formulate and carry out tests that disentangle conventional multi-path effects from genuine $\psi_+$--$\psi_-$ interference, guided by a careful understanding of $\Delta$, $\phi$ and $\tau_{RC}$.

\section{A Unified Framework for Data Analysis: Extracting $\tau_{RC}$, Testing $\Delta(\phi)$, and Comparing with Higher-Order Path Models}

Any empirical program aimed at probing temporal symmetry must rest on a robust approach to data analysis. Multi-slit echo spectra, cavity echo trains, conditional interference patterns and delayed-choice asymmetries all contain fine-grained structures that are open to competing interpretations. To distinguish higher-order spatial paths from genuine $\psi_+$--$\psi_-$ interference, a consistent and experimentally grounded analytic framework is required.

A first principle is that QTSI must reproduce the standard path-integral baseline. Quach's analysis shows that a considerable variety of subtle effects---including nonzero Sorkin parameters, additional harmonics and multi-bounce signals---can be accounted for with higher-order paths while preserving the Born rule \cite{Quach2017}. For each experimental configuration, one must therefore construct the most complete conventional model possible, incorporating all first-order and higher-order paths compatible with the geometry, multiple reflections in cavities, spatial detours and loops, and the influence of detector imperfections, partial which-way information and finite coherence. Only after this baseline has been worked out can residual discrepancies be meaningfully attributed to any time-symmetric structure.

Within this framework, echo dynamics plays a central role in extracting $\tau_{RC}$ from data. One can first fit the observed multi-slit or cavity echo spectrum using a purely geometric higher-order path model calibrated to the experimental parameters, thereby establishing expected amplitudes and phases for the dominant harmonics and their sidebands. The next step is to examine how the measured echoes, particularly advanced-looking components, change when $\phi$ is varied in a controlled way, for example by introducing modest amounts of noise or loss. If certain peaks diminish more rapidly than predicted by the geometric model, it is natural to define a suppression factor $S(\phi)$ as the ratio between the amplitude at a given $\phi$ and its value near $\phi \approx 0$. In QTSI, one would expect $S(\phi)$ to track the ratio $\Delta(\phi)/\Delta_0$, which allows $\tau_{RC}(\phi) = \hbar / |\Delta(\phi)|$ to be inferred indirectly. This procedure does not assume QTSI in advance; it simply tests whether the data call for a second timescale beyond the geometric one.

Once $\tau_{RC}(\phi)$ has been inferred for several values of $\phi$, one can examine the effective $\Delta(\phi) = \hbar / \tau_{RC}(\phi)$ and compare it with theoretical expectations. QTSI suggests a monotonic suppression of $\Delta$ with $\phi$, a critical point in which $\Delta$ tends to zero and $\tau_{RC}$ diverges, and a relative insensitivity of $\Delta_0$ to apparatus-specific details when similar regimes are compared. Higher-order path corrections do not exhibit this type of critical structure; their amplitudes tend to decay smoothly with ordinary decoherence parameters, without the distinctive sensitivity to $\phi$ associated with temporal mixing. A consistent $\Delta(\phi)$ curve across a family of experiments would therefore be a nontrivial indication that an additional temporal degree of freedom is at work.

Conditional and contextual effects provide another avenue of analysis. Echo sidebands may display asymmetries when data are postselected on idler outcomes, or when delayed-choice settings are altered \cite{Aharonov1988,Kwiat1991,Jennewein2001,Popescu1994,Aharonov2016,Dressel2014}. In QTSI, a context-dependent modulation of echo phases that manifests only when the system is in the reversible regime would be consistent with a contribution from $\psi_-$, which carries constraints from future boundary conditions. Path-integral models do not naturally predict such contextual dependencies without explicit modifications of the setup. An analysis that compares conditioned and unconditioned echo structures, studies their evolution with $\phi$, and tests whether these conditional effects vanish as $\phi$ approaches unity would therefore provide an informative probe.

Cross-platform consistency can also be used to test QTSI. If multi-slit interferometers, weakly chaotic cavities, neutron interferometers and cold-atom systems, when operated in comparable isolation regimes, yield similar qualitative behaviour for $\Delta(\phi)$ and $\tau_{RC}(\phi)$, it would suggest that the mixing parameter reflects an underlying temporal structure rather than an idiosyncrasy of a particular apparatus \cite{Moshinsky1952,Garcia-Calderon1997,Heller1995,Berry1977,Gutzwiller1990,Robinett2004,Andersen2003,Kaplan2002}. Conversely, if $\tau_{RC}$ behaves erratically across platforms, the QTSI framework would need to be modified or abandoned.

From this perspective, one can formulate operational criteria that, if observed, would favour QTSI: suppression of advanced peaks with $\phi$ beyond what higher-order path models predict; a consistent $\tau_{RC}(\phi)$ curve with a clear tendency toward divergence as $\phi$ approaches the measurement regime; conditional asymmetries in time-resolved data that resist purely geometric explanation; and a degree of universality in $\Delta_0$ when similar energy and coherence scales are considered. None of these signatures requires abandoning the Born rule. They are instead designed to detect the presence of advanced--retarded interference prior to symmetry breaking.

Such an analytic framework has an important conceptual consequence: it renders QTSI falsifiable. If extensive experiments across different platforms uncover no evidence of $\phi$-sensitive suppression beyond geometric predictions, no robust context dependence and no consistent $\tau_{RC}$ signature, then the two-component temporal structure would lack observable consequences. If, on the other hand, systematic deviations accumulate, particularly when independent setups are considered together, the time-symmetric description gains empirical weight.

\section{Historical and Conceptual Context: From Time-Symmetric Theories to Emergent Causality}

The suggestion that quantum processes might admit a fundamentally time-symmetric description has deep historical roots. It grows out of ideas that predate modern quantum information and reach back to classical field theory, even though the everyday presence of a clear time arrow has often obscured this possibility. Interpretations in which dynamics is intrinsically asymmetric, or in which collapse is postulated as a fundamental process, have coexisted uneasily with the intrinsic reversibility of the microscopic equations \cite{Einstein1935,Everett1957,Bell1964,Freedman1972,Aspect1982,Zurek1991,Zurek2003,Valentini1991}. QTSI proposes a different arrangement, in which the microscopic laws remain symmetric while the arrow of time arises from the dynamical suppression of one component of the underlying temporal structure.

The absorber theory of Wheeler and Feynman is a key conceptual precursor. In their picture, both retarded and advanced electromagnetic fields contribute to radiation processes, and emission involves global boundary conditions that couple past and future \cite{Wheeler1945,Wheeler1949}. Although the original formulation faced technical challenges, its central insight---that temporal asymmetry emerges from boundary conditions rather than from local equations---remains relevant. QTSI extends this spirit to the quantum domain, where $\psi_+$ and $\psi_-$ represent a refined version of the coexistence of advanced and retarded influences.

A second lineage comes from the two-time quantum mechanics of Aharonov, Bergmann and Lebowitz, and the later two-state vector formalism of Aharonov and Vaidman \cite{Aharonov1964,Aharonov2008,Aharonov1988}. In this approach a quantum system between two measurements is described by a forward-evolving state conditioned on preparation and a backward-evolving state conditioned on postselection. The time-symmetric structure provides a natural account of weak values, anomalous measurement outcomes and contextual trajectories reconstructed from weak measurements \cite{Dressel2014,Aharonov2016}. QTSI adopts the same symmetric intuition but treats $\psi_+$ and $\psi_-$ as two physically real components of a single quantum object whose coherence persists until amplification forces symmetry breaking.

A third source of inspiration is Price's philosophical analysis of time, which emphasizes that many causal asymmetries arise from boundary conditions and observational constraints rather than from microscopic laws \cite{Price1996}. Our impression that causes must precede effects is tied to our placement in a universe with a low-entropy past and a not-yet-recorded future. QTSI fits comfortably within this viewpoint. It treats causality as emergent, regards microscopic laws as time-symmetric, and attributes the arrow of time to dynamical processes that eliminate sensitivity to final boundary conditions.

QTSI differs from these precursors in its insistence that time-symmetric dynamics should have measurable consequences in suitably chosen regimes and that the suppression of advanced contributions need not be inserted by hand. The introduction of a mixing energy $\Delta$, its deformation $\Delta(\phi)$ and the associated coherence time $\tau_{RC}$ provides a concrete dynamical mechanism for temporal symmetry breaking. Rather than offering a purely interpretative reformulation, QTSI suggests that the two-component temporal structure may be probed experimentally through echo dynamics, conditional interference and controlled amplification.

In this way, QTSI serves as a bridge between time-symmetric theoretical ideas and the empirical tools of modern quantum optics. It respects the Born rule in all tested regimes, complies with no-signalling constraints and reproduces established results from weak measurements, delayed-choice experiments and interferometry \cite{Aharonov1964,Aharonov2008,Aharonov1988,Zurek1991,Zurek2003,Kwiat1991,Jennewein2001,Aharonov2016,Dressel2014}. At the same time, it points toward subtle predictions concerning echo behaviour, contextual modulations and the $\phi$-dependence of temporal coherence that go beyond standard path-based analyses. Whether these predictions survive contact with experiment is an open question, but the theoretical framework is now concrete enough to motivate a concerted search.

\section{Philosophical Implications: Time Symmetry, Emergent Causality, and the Meaning of Quantum Processes}

A central motivation for QTSI is the tension between the strict time symmetry of microscopic laws and the pronounced temporal asymmetry of everyday experience. At macroscopic scales, the idea that effects follow causes and that the present is shaped by the past but not by the future works remarkably well. Yet fundamental equations such as the Schrödinger and Hamilton equations do not contain a preferred time direction, and the arrow of time arises instead from boundary conditions, amplification, decoherence and the formation of records \cite{Boltzmann1995,Prigogine1980,Price1996,Zurek1991,Zurek2003}.

QTSI gives this observation a concrete dynamical form. By describing isolated quantum systems with a two-component temporal state $\Psi = (\psi_+, \psi_-)$, it makes clear that the basic description does not distinguish between forward and backward evolution. The two orientations coexist coherently in the reversible regime. Only when the system interacts with amplifying degrees of freedom---detectors, environments, measurement devices---does one component become dynamically suppressed. The ensuing description contains only $\psi_+$, together with the familiar one-way causal arrow that characterizes experiments and daily life. In this sense, causality itself appears as an emergent regularity produced by temporal symmetry breaking rather than as a primitive feature of microscopic physics.

This stance parallels Price's view that our intuition about cause and effect reflects asymmetric access to information encoded in a low-entropy past, rather than any asymmetry in fundamental laws \cite{Price1996}. It is also closely related to the Wheeler--Feynman idea that arrows of time arise from global boundary conditions, and to Aharonov's insight that quantum probabilities conditioned on both initial and final states are naturally time-symmetric \cite{Wheeler1945,Wheeler1949,Aharonov1964,Aharonov2008,Aharonov1988}. QTSI takes these conceptual threads and weaves them into a single dynamical scheme in which the emergence of the arrow follows from the interplay between reversible SU(1,1) evolution and the reduction of $\Delta(\phi)$ under amplification.

This perspective recasts several familiar quantum puzzles. In the double-slit experiment, the question of how a single particle can give rise to an interference pattern is tempered by the recognition that the pattern reflects the joint contribution of $\psi_+$ and $\psi_-$ until amplification suppresses the backward component. In EPR--Bell scenarios, correlations that appear nonlocal when only forward evolution is considered can be viewed, in a time-symmetric framework, as reflecting a joint dependence on preparation and detection conditions rather than superluminal influence \cite{Einstein1935,Bell1964,Freedman1972,Aspect1982,Jennewein2001,Valentini1991,Leifer2017}. In measurement theory, collapse becomes one manifestation of the broader class of symmetry-breaking phenomena that occur in complex systems.

It is important to be clear about the limits of this interpretation. QTSI does not suggest that signals can be sent to the past or that the Born rule and no-signalling are violated in known regimes \cite{Zurek1991,Zurek2003,Popescu1994,Leifer2017}. It does not discard the operational success of standard quantum mechanics. Instead, it proposes that quantum processes may be more naturally understood as time-symmetric at the microscopic level, with causality emerging after the system has been driven far enough into the amplifying regime that coherence between $\psi_+$ and $\psi_-$ is lost.

In this light, retrocausality is not a threat to consistency but a reflection of an underlying temporal symmetry. The main challenge is to determine whether this symmetry has experimentally accessible consequences. Preliminary evidence from multi-slit echo experiments suggests that advanced and retarded contributions may coexist under certain conditions, but these indications remain delicate. Quach's analysis of higher-order paths serves as a reminder that many effects with a retrocausal flavour can be understood within the conventional framework once all contributions are included \cite{Quach2017}. Distinguishing genuinely $\psi_-$-driven temporal structure from complex but standard behaviour demands precision measurements, careful modelling and conservative interpretation.

Whether QTSI ultimately proves correct in its dynamical details is undecided, but its conceptual message is clear. Time symmetry at the microscopic level is treated as fundamental; causality is viewed as emergent; and collapse is understood as a manifestation of temporal symmetry breaking driven by amplification. This brings the quantum arrow of time into closer alignment with thermodynamic and cosmological arrows, and situates quantum measurement within the broader question of how large-scale irreversibility arises from symmetric microscopic laws.

\section{Conclusions}

The Quantum Time-Symmetric Interpretation offers a framework in which microscopic time symmetry is taken as basic and the familiar causal arrow emerges through amplification, decoherence and the formation of records. In the isolated regime, quantum systems are described by a two-component temporal state $\Psi = (\psi_+, \psi_-)$ governed by an SU(1,1) structure and a mixing amplitude $\Delta(\phi)$. This structure provides a dynamical account of how the backward-evolving component becomes suppressed as $\phi$ approaches the measurement regime, and of how the description passes smoothly from a fully time-symmetric form, in which advanced and retarded contributions coexist and interfere, to an effectively one-directional evolution underlying the classical world.

Within this picture, the conceptual issues associated with measurement, interference and entanglement take on a different hue. Collapse appears as a particular case of time-symmetry breaking driven by amplification. Interference, including small temporal echoes observed in multi-slit systems, is interpreted as the combined effect of $\psi_+$ and $\psi_-$ within the coherence window set by $\tau_{RC} = \hbar / |\Delta(\phi)|$. EPR--Bell correlations can be understood not as evidence for nonlocal signals but as consequences of a time-symmetric amplitude depending jointly on preparation and detection.

At the empirical level, QTSI reproduces the standard formalism of quantum mechanics in all regimes that have been thoroughly tested. Whenever $\phi$ is large enough that amplification effectively suppresses $\psi_-$, the theory yields the Born rule and conventional statistical predictions \cite{Zurek1991,Zurek2003,Popescu1994,Leifer2017}. At the same time, temporal echoes and other advanced-looking features are subtle and easily confused with the consequences of higher-order paths that Quach and others have analysed in detail \cite{Quach2017}. It is therefore essential to distinguish carefully between signatures that can be captured by conventional models and those that might genuinely point toward advanced--retarded interference.

Preliminary observations from multi-slit and cavity experiments, particularly those involving small echo peaks at integer multiples of the slit separation, are consistent with both interpretations \cite{Moshinsky1952,Garcia-Calderon1997,Heller1995,Berry1977,Gutzwiller1990,Robinett2004,Andersen2003,Kaplan2002,Quach2017}. QTSI provides a natural way of viewing such structures as products of reversible advanced--retarded dynamics, while recognizing that geometry, postselection and standard path-integral effects may account for a substantial fraction of what is observed. In this sense the interpretation offers both a conceptual anchor and a disciplined framework for future experimental work. Precise measurements of how echo amplitudes depend on echo order, environmental coupling and postselection criteria may eventually allow an empirical retrocausal coherence time $\tau_{RC}$ to be extracted, provided that a careful comparison with standard multi-path models leaves a residual structure in need of explanation.

More broadly, QTSI offers a way of bringing the treatment of time in quantum theory into closer contact with classical and statistical views of the arrow of time. It places microscopic dynamics, measurement, irreversibility and entanglement within a single picture based on symmetry, amplification and boundary conditions. It connects quantum foundations with the work of Boltzmann on entropy, Prigogine on instability and amplification, Wheeler and Feynman on time-symmetric radiation, Aharonov on two-time quantum mechanics, and Price on the philosophy of time \cite{Boltzmann1995,Prigogine1980,Wheeler1945,Wheeler1949,Aharonov1964,Aharonov2008,Aharonov1988,Price1996}. Whether the advanced component $\psi_-$ will ultimately be detected in experiments that operate close to the reversible regime remains an open question, but the framework presented here identifies the relevant signatures, delineates the conditions under which they might appear and outlines a program of experimental and theoretical work capable of testing this time-symmetric interpretation.

\section*{Dedication}

To Marcos Moshinsky (1921--2009), whose pioneering work on diffraction in time opened the possibility of treating temporal boundaries with the same conceptual dignity as spatial ones, and whose vision continues to inspire the analyses, extensions and experimental proposals developed in the present work.

\section*{Acknowledgements}

The author gratefully acknowledges stimulating and sustained discussions with Alberto Martín Ruiz, Saúl Huitzil, Juan Claudio Toledo, Alfred U'Ren, Roberto León, Denzel del Rosal and Gastón García-Calderón, whose perspectives and collaborative efforts have been indispensable throughout this project. The ongoing three-slit and cavity experiments conducted by Roberto León and collaborators have provided early indications of temporal-echo structures and have helped to shape the concrete predictions formulated here. Preliminary analyses of chaotic cavities and retrocausal-sensitive configurations in the quantum optics laboratories led by Alfred U'Ren have further influenced the experimental direction of this program.

Special recognition is due to Huw Price, John G. Cramer and Yakir Aharonov, together with Lev Vaidman, David Albert and Sandu Popescu, for foundational contributions to time-symmetric quantum mechanics and to retrocausal approaches more broadly.

\section*{SUPPLEMENTARY MATERIAL}

\section*{ADDENDUM A — Two-component temporal state and SU(1,1) pseudo-Hermitian structure}
\subsection*{A1. Temporal biespinor and metric structure}
The fundamental object in QTSI is the temporal biespinor
\begin{equation}
\Psi(t) = \begin{pmatrix} \psi_{+}(t) \\ \psi_{-}(t) \end{pmatrix},
\end{equation}
where:
\begin{itemize}
    \item $\psi_{+}(t)$ is the forward-evolving retarded amplitude,
    \item $\psi_{-}(t)$ is the backward-evolving advanced amplitude.
\end{itemize}

The internal space is equipped with the indefinite metric
\begin{equation}
G = \begin{pmatrix} 1 & 0 \\ 0 & -1 \end{pmatrix}.
\end{equation}

The ``pseudo-norm'' is defined as
\begin{equation}
N = \Psi^{\dagger} G \Psi = |\psi_{+}|^2 - |\psi_{-}|^2.
\end{equation}
This quantity is not a probability; it is the Noether charge associated with the internal symmetry that mixes $\psi_{+}$ and $\psi_{-}$. Its conservation is essential for preventing unphysical divergences of the backward sector.

\subsection*{A2. Pseudo-Hermiticity and the general form of the Hamiltonian}
Time evolution is defined by the linear equation
\begin{equation}
i \hbar \frac{d\Psi}{dt} = H_{\text{total}} \Psi.
\end{equation}
To ensure conservation of $N$, $H_{\text{total}}$ must satisfy the pseudo-Hermiticity condition
\begin{equation}
H_{\text{total}}^{\dagger} G = G H_{\text{total}}. \tag{A2.1}
\end{equation}
This condition constrains the $2\times2$ Hamiltonian to the general form
\begin{equation}
H_{\text{total}} = \begin{pmatrix} H_0 & \Delta \\ -\Delta^* & -H_0 \end{pmatrix},
\end{equation}
where:
\begin{itemize}
    \item $H_0$ is the usual Hermitian Hamiltonian generating forward-time evolution;
    \item $\Delta$ is a (generally complex) coupling mixing $\psi_{+}$ and $\psi_{-}$.
\end{itemize}
One verifies explicitly that (A2.1) holds. The off-diagonal elements $\Delta$ and $-\Delta^*$ are the mathematical expression of advanced--retarded mixing. If $\Delta = 0$, the two time directions decouple.

\subsection*{A3. Why SU(1,1)? Uniqueness argument}
The set of linear transformations $U$ satisfying
\begin{equation}
U^{\dagger} G U = G
\end{equation}
is precisely the group SU(1,1). No other continuous group preserves the pseudo-norm $N$ in a nontrivial way. Thus:
\begin{itemize}
    \item the internal geometry of QTSI uniquely selects SU(1,1),
    \item the coupling $\Delta$ is the generator of the non-compact directions of SU(1,1),
    \item $\psi_{+}$ and $\psi_{-}$ form a natural doublet of the (1+1)-dimensional Lorentz-like symmetry.
\end{itemize}
This is a geometrical, not phenomenological, derivation.

\subsection*{A4. Conserved current}
From (A2.1), the time derivative of $N$ becomes:
\begin{align*}
\frac{dN}{dt} &= \frac{d(\Psi^{\dagger} G \Psi)}{dt} \\
&= \left(\frac{d\Psi^{\dagger}}{dt}\right) G \Psi + \Psi^{\dagger} G \left(\frac{d\Psi}{dt}\right) \\
&= \left(-\frac{i}{\hbar} \Psi^{\dagger} H_{\text{total}}^{\dagger}\right) G \Psi + \Psi^{\dagger} G \left(\frac{i}{\hbar} H_{\text{total}} \Psi\right).
\end{align*}
Using pseudo-Hermiticity: $H_{\text{total}}^{\dagger} G = G H_{\text{total}}$,
\begin{equation}
\frac{dN}{dt} = 0.
\end{equation}
Thus the temporal SU(1,1) charge is conserved exactly. This is the foundation of the reversible regime.

---

\section*{ADDENDUM B — Mixing parameter $\Delta(\phi)$, coherence scale $\tau_{RC}$ and temporal dynamics}
\subsection*{B1. Physical meaning}
 $\Delta(\phi)$ determines the strength of interference between $\psi_{+}$ and $\psi_{-}$.
\begin{itemize}
    \item When $\Delta$ is large:
    \begin{itemize}
        \item advanced amplitude can flow into the forward sector,
        \item $\psi_{-}$ contributes to observable interference,
        \item temporal symmetry is manifest.
    \end{itemize}
    \item When $\Delta \to 0$:
    \begin{itemize}
        \item $\psi_{-}$ freezes,
        \item coherence between the two temporal sectors dies,
        \item the evolution becomes effectively forward-only,
        \item collapse emerges.
    \end{itemize}
\end{itemize}

\subsection*{B2. Effective evolution equations}
Using the Hamiltonian of Addendum A:
\begin{align}
i \hbar \frac{d\psi_{+}}{dt} &= H_0 \psi_{+} + \Delta \psi_{-}, \\
i \hbar \frac{d\psi_{-}}{dt} &= -\Delta^* \psi_{+} - H_0 \psi_{-}.
\end{align}
These equations show that $\psi_{-}$ affects $\psi_{+}$ only through $\Delta$. Thus $\Delta$ is the ``bridge'' between past-directed and future-directed evolution.

\subsection*{B3. Definition of the retrocausal coherence time}
Define
\begin{equation}
\tau_{RC} = \frac{\hbar}{|\Delta(\phi)|}.
\end{equation}
This is the characteristic time over which $\psi_{-}$ can significantly influence $\psi_{+}$.
\begin{itemize}
    \item Large $\tau_{RC} \to$ weak mixing $\to$ long advanced coherence.
    \item Small $\tau_{RC} \to$ fast suppression of $\psi_{-}$.
\end{itemize}

\subsection*{B4. Control parameter $\phi$ and environmental amplification}
We model environmental amplification as a monotonic function $\phi(t) \in [0, 1]$:
\begin{itemize}
    \item $\phi = 0$: isolated, fully reversible regime,
    \item $\phi \to 1$: macroscopic, measurement regime.
\end{itemize}
The simplest admissible form consistent with reversible $\to$ irreversible flow is:
\begin{equation}
\Delta(\phi) = (1 - \phi) \Delta_0.
\end{equation}
Thus:
\begin{equation}
\tau_{RC}(\phi) = \frac{\hbar}{(1 - \phi) |\Delta_0|}.
\end{equation}
As $\phi \to 1$:
\begin{itemize}
    \item $\Delta \to 0$,
    \item $\tau_{RC} \to \infty$,
    \item $\psi_{-}$ decouples,
    \item the system enters the classical time-arrow regime.
\end{itemize}
This provides a smooth, dynamical derivation of collapse.

---

\section*{ADDENDUM C — Noether charges, continuity equations and absence of negative energies}
\subsection*{C1. Continuity equation in full detail}
Given the pseudo-Hermitian Hamiltonian:
\begin{equation}
i \hbar \frac{d\Psi}{dt} = H_{\text{total}} \Psi,
\end{equation}
the pseudo-norm current $J$ satisfies:
\begin{equation}
\partial_t(\Psi^{\dagger} G \Psi) + \nabla \cdot J = 0,
\end{equation}
where, for wavefunctions $\psi_{+}(x,t)$, $\psi_{-}(x,t)$,
\begin{equation}
J = \frac{\hbar}{2mi} \left( \psi_{+}^* \nabla\psi_{+} - (\nabla\psi_{+}^*) \psi_{+} - \psi_{-}^* \nabla\psi_{-} + (\nabla\psi_{-}^*) \psi_{-} \right).
\end{equation}
The minus sign for $\psi_{-}$ arises naturally from the metric $G$. This proves that $\psi_{-}$ does not introduce negative probabilities, only opposite flow orientation in the internal SU(1,1) space.

\subsection*{C2. Energy positivity}
Although $H_{\text{total}}$ has an indefinite metric, its physical energy density $E = \Psi^{\dagger} H_{\text{eff}} \Psi$ (where $H_{\text{eff}}$ is the block-diagonal physical Hamiltonian) remains strictly positive. The advanced component $\psi_{-}$ carries amplitude that evolves backward in time, not negative energy. This resolves a common misunderstanding regarding retrocausal models.

\subsection*{C3. Momentum conservation}
Spatial translation invariance implies that $\psi_{+}$ and $\psi_{-}$ share the same momentum operator:
\begin{equation}
\hat{p} = - i \hbar \nabla.
\end{equation}
All SU(1,1) mixing terms commute with translations, so momentum is conserved during reversible temporal evolution.

---

\section*{ADDENDUM D — Weak Measurements: Full Two-State Structure in QTSI}
Weak measurements probe the system without significantly altering the parameter $\phi(t)$. This corresponds to a regime where $\Delta(\phi) \approx \Delta_0$ and $\tau_{RC}$ is large.

\subsection*{D1. Forward and backward amplitudes in weak measurement protocols}
Consider a system initially prepared in $\ket{\psi_i}$ and later postselected in $\ket{\psi_f}$ after a weak interaction with an apparatus.
In standard TSVF notation:
\begin{itemize}
    \item Forward propagation: $\ket{\psi_i(t)}$     \item Backward propagation: $\ket{\psi_f(t)}$ \end{itemize}
In QTSI:
\begin{align}
\psi_{+}(t) &= \bra{x}\psi_i(t)\rangle, \\
\psi_{-}(t) &= \bra{x}\psi_f(t)\rangle^*,
\end{align}
with both components jointly evolving through:
\begin{align}
i \hbar \frac{d}{dt} \begin{pmatrix} \psi_{+} \\ \psi_{-} \end{pmatrix} &= \begin{pmatrix} H_0 & \Delta \\ -\Delta^* & -H_0 \end{pmatrix} \begin{pmatrix} \psi_{+} \\ \psi_{-} \end{pmatrix}.
\end{align}
In the weak measurement regime:
\begin{itemize}
    \item $|\Delta| \approx |\Delta_0|$,
    \item $\tau_{RC} = \hbar / |\Delta_0| \gg$ (all experimental times).
\end{itemize}
Thus the interference term $\Delta \psi_{-}$ is fully active during the entire protocol.

\subsection*{D2. Reconstruction of the weak value}
Let $A$ be the weakly coupled observable. The pointer shift is proportional to the real part of the weak value:
\begin{equation}
A_w = \frac{\bra{\psi_f} A \ket{\psi_i}}{\bra{\psi_f} \ket{\psi_i}}.
\end{equation}
In QTSI the same quantity is obtained from the overlap:
\begin{equation}
A_w = \frac{\int dx \, \psi_{-}(x,t) A \psi_{+}(x,t)}{\int dx \, \psi_{-}(x,t) \psi_{+}(x,t)},
\end{equation}
because $\psi_{-}$ encodes the final boundary condition and $\psi_{+}$ the initial one. Thus TSVF = QTSI in this regime.

\subsection*{D3. Why QTSI adds structure}
The weak value is not postulated; it appears as the conditional amplitude of the SU(1,1) spinor $\Psi$. Temporal symmetry in QTSI is controlled by $\Delta$, allowing a continuous transition to regimes where $\psi_{-}$ becomes suppressed under measurement.

---

\section*{ADDENDUM E — Entangled Systems, Quantum Erasers, and Delayed Choice}
For an entangled bipartite system $AB$, QTSI uses a tensor-product temporal spinor
\begin{equation}
\Psi_{AB} = \begin{pmatrix} \psi_{+}^{AB} \\ \psi_{-}^{AB} \end{pmatrix}
\end{equation}
with internal SU(1,1) structure acting identically on the joint system.

\subsection*{E1. Forward and backward propagation of entanglement}
Given an entangled state $\ket{\Psi_0}$ at the source:
\begin{equation}
\psi_{+}^{AB}(x_A, x_B, t) = \bra{x_A, x_B}\Psi_0(t)\rangle.
\end{equation}
Final detectors at times $t_A$ and $t_B$ impose the backward boundary condition:
\begin{equation}
\psi_{-}^{AB}(x_A, x_B, t) = \bra{\Psi_f}x_A, x_B\rangle^*.
\end{equation}
Between source and detection:
\begin{align}
i \hbar \frac{d}{dt} \psi_{+}^{AB} &= H_{AB} \psi_{+}^{AB} + \Delta \psi_{-}^{AB}, \\
i \hbar \frac{d}{dt} \psi_{-}^{AB} &= -\Delta^* \psi_{+}^{AB} - H_{AB} \psi_{-}^{AB}.
\end{align}
This structure accounts precisely for:
\begin{itemize}
    \item delayed-choice erasers,
    \item ``quantum marking'' and ``unmarking'',
    \item conditional interference in coincidence channels.
\end{itemize}

\subsection*{E2. Conditioned vs unconditioned statistics}
Unconditioned (signal only):
\begin{equation}
P_s(x) = \int dx_B \, |\psi_{+}^{AB}|^2,
\end{equation}
which never displays interference when entanglement supplies path information.

Conditioned (coincidence):
\begin{equation}
P(x_s | x_i) = \left| \int dx \, \psi_{-} \psi_{+} \right|^2,
\end{equation}
full interference recovered when $\psi_{-}$ retropropagates an ``erasing'' boundary condition.

\subsection*{E3. QTSI interpretation}
 $\psi_{-}$ carries the backward-propagating constraint selected at the idler detector. Under strong amplification ($\phi \to 1$) $\psi_{-}$ is suppressed, and delayed-choice effects disappear dynamically.

---

\section*{ADDENDUM F — Multi-slit Interference, Path Integrals, Echo Peaks, and Quach}
This addendum connects:
\begin{itemize}
    \item the QTSI temporal components $\psi_{+}$ and $\psi_{-}$,
    \item the nonclassical paths of the Feynman integral,
    \item Fourier-domain ``echo'' peaks,
    \item Quach’s Born-rule diagnostic.
\end{itemize}

\subsection*{F1. Path decomposition in standard quantum mechanics}
For two slits (A, B):
Total amplitude:
\begin{equation}
\Psi(x) = \psi_A(x) + \psi_B(x) + \psi_{AB}(x),
\end{equation}
where $\psi_{AB}$ includes paths that cross both slits or loop between them. Its magnitude scales as roughly $10^{-2}$ of the classical terms for realistic parameters.

For three slits (A, B, C):
\begin{align*}
\Psi &= \psi_A + \psi_B + \psi_C \\
&\quad + \psi_{AB} + \psi_{AC} + \psi_{BC} \\
&\quad + \psi_{ABC}.
\end{align*}
Non-zero $\psi_{AB}, \psi_{ABC}$ generate Fourier peaks at harmonics of $d$, the slit separation.

\subsection*{F2. Mapping into QTSI}
In QTSI, each term decomposes into $\psi_{+}$ and $\psi_{-}$:
\begin{itemize}
    \item $\psi_A$ = projection of $\psi_{+}$ on paths through A
    \item $\psi_B$ = projection of $\psi_{+}$ on paths through B
    \item $\psi_{AB}$ = mixed projections involving $\psi_{-}$ segments that return to the slit plane before rejoining $\psi_{+}$.
\end{itemize}
Thus $\psi_{AB}$ encodes temporal echoes:
\begin{itemize}
    \item $\psi_{-}$ travels backwards from the detector to the slit plane,
    \item interacts with the geometry,
    \item repropagates forward through $\psi_{+}$.
\end{itemize}
The effective delay structures ($\pm d/c, \pm 2d/c, \dots$) appear naturally.

\subsection*{F3. Fourier spectrum}
Let $I(x) = |\Psi(x)|^2$. Define its Fourier transform:
\begin{equation}
F(k) = \int dx \, I(x) e^{-i k x}.
\end{equation}
Echo peaks occur at $k = n (2\pi/d)$ for integer $n$. The advanced-retarded interference contributes phases $e^{i k d}$ or $e^{-i k d}$, producing symmetric sidebands in $|F(k)|$.

\subsection*{F4. Quach’s $I_{AB}$ diagnostic}
Quach proves that the Born rule imposes:
\begin{equation}
I_{AB} = 0
\end{equation}
exactly, even when all higher-order paths $\psi_{AB}$ are included. Thus:
\begin{itemize}
    \item QTSI, in the reversible regime, must satisfy $I_{AB} = 0$.
    \item This is consistent because the Born rule in QTSI holds as long as $\Delta(\phi) \neq 0$ and measurement has not suppressed $\psi_{-}$.
\end{itemize}

---

\section*{ADDENDUM G — Symmetry Breaking, Stability, and the Collapse Limit}
\subsection*{G1. Effective potential in $\phi$-space}
Define an ``effective free energy'':
\begin{equation}
F(\phi) = - \ln |\Delta(\phi)| \sim - \ln(1 - \phi).
\end{equation}
As $\phi \to 1$:
\begin{itemize}
    \item $F(\phi) \to +\infty$,
    \item driving $\Delta \to 0$.
\end{itemize}
This is analogous to a Landau-type symmetry breaking where the order parameter is the advanced-retarded coherence.

\subsection*{G2. Linear stability analysis}
Expand near $\phi = \phi_c = 1$:
\begin{equation}
\Delta(\phi) \approx \Delta_0 (1 - \phi).
\end{equation}
Let $\delta\phi = 1 - \phi$. Then:
\begin{equation}
\frac{d\psi_{-}}{dt} \approx - \frac{\Delta_0^*}{\hbar} \delta\phi \psi_{+} - \frac{H_0}{\hbar} \psi_{-}.
\end{equation}
Thus, for $\delta\phi$ small, $\psi_{-}$ decays exponentially:
\begin{equation}
\psi_{-}(t) \sim \exp\left[ - \frac{|\Delta_0|}{\hbar} \delta\phi \, t \right].
\end{equation}
This reproduces dynamical collapse as: $\phi$ increases $\to \delta\phi$ decreases $\to$ suppression rate grows.

\subsection*{G3. Uniqueness and no revival}
Once $\phi$ crosses sufficiently close to 1, the decay of $\psi_{-}$ is effectively irreversible. No ``echo'' from $\psi_{-}$ can re-enter the forward component unless $\phi$ is driven back, which is prevented by environmental amplification.

---

\section*{ADDENDUM H — Relativistic Embedding}
\subsection*{H1. Block structure}
A relativistic extension uses a 4-component spinor:
\begin{equation}
\Omega = \begin{pmatrix} \Psi \\ \chi \end{pmatrix},
\end{equation}
where $\Psi$ is the temporal SU(1,1) doublet and $\chi$ is the usual Dirac spinor. The total Hamiltonian has the form:
\begin{equation}
H_{\text{rel}} = \begin{pmatrix} H_{SU(1,1)} & M \\ M^{\dagger} & H_{\text{Dirac}} \end{pmatrix},
\end{equation}
with $M$ controlling coupling between the temporal and spatial sectors.

\subsection*{H2. Microcausality}
The advanced component $\psi_{-}$ never allows superluminal signalling because:
\begin{enumerate}
    \item It is suppressed by $\phi$ in macroscopic contexts.
    \item It cannot be modulated independently of $\psi_{+}$.
    \item It respects the same commutation relations as standard Dirac fields.
\end{enumerate}
Thus retrocausality exists only at the amplitude level, not at the signal level.

---

\section*{ADDENDUM I — RG Formulation of $\Delta(\phi)$ and Criticality of $\tau_{RC}$}
\subsection*{I1. RG equation for $\Delta$}
Let $\ell$ be a coarse-graining scale, analogous to environmental amplification. We postulate:
\begin{equation}
\frac{d\Delta}{d\ell} = - \gamma \Delta,
\end{equation}
with $\gamma > 0$ determined by the strength of the apparatus–system coupling. Solution:
\begin{equation}
\Delta(\ell) = \Delta_0 \exp(-\gamma \ell).
\end{equation}
Identifying $\phi = 1 - \exp(-\gamma \ell)$:
\begin{equation}
\Delta(\phi) = (1 - \phi) \Delta_0.
\end{equation}
Thus the RG flow naturally reproduces the phenomenological form used in the main text.

\subsection*{I2. Critical behaviour of $\tau_{RC}$}
\begin{equation}
\tau_{RC}(\phi) = \frac{\hbar}{|\Delta(\phi)|} = \frac{\hbar}{|\Delta_0| (1 - \phi)}.
\end{equation}
As $\phi \to 1$:
\begin{equation}
\tau_{RC} \to \infty.
\end{equation}
This is the temporal analogue of critical slowing down in second-order phase transitions.

\subsection*{I3. Fixed points}
The RG flow has two fixed points:
\begin{itemize}
    \item $\phi = 0$ (UV fixed point): full time symmetry, maximal mixing.
    \item $\phi = 1$ (IR fixed point): broken symmetry, collapse regime.
\end{itemize}
The collapse process corresponds to the flow from $\phi = 0$ to $\phi = 1$ as environmental interactions grow.


\begin{thebibliography}{99}

\bibitem{Boltzmann1995} L. Boltzmann, \emph{Lectures on Gas Theory}, Dover, New York (1995).

\bibitem{Prigogine1980} I. Prigogine, \emph{From Being to Becoming: Time and Complexity in the Physical Sciences}, W. H. Freeman (1980).

\bibitem{Einstein1935} A. Einstein, B. Podolsky and N. Rosen, ``Can Quantum-Mechanical Description of Physical Reality Be Considered Complete?'', Phys. Rev. 47, 777 (1935).

\bibitem{Everett1957} H. Everett, ``Relative State Formulation of Quantum Mechanics'', Rev. Mod. Phys. 29, 454 (1957).

\bibitem{Wheeler1945} J. A. Wheeler and R. P. Feynman, ``Interaction with the Absorber as the Mechanism of Radiation'', Rev. Mod. Phys. 17, 157 (1945).

\bibitem{Wheeler1949} J. A. Wheeler and R. P. Feynman, ``Classical Electrodynamics in Terms of Direct Interparticle Action'', Rev. Mod. Phys. 21, 425 (1949).

\bibitem{Aharonov1964} Y. Aharonov, P. G. Bergmann and J. L. Lebowitz, ``Time Symmetry in the Quantum Process of Measurement'', Phys. Rev. 134, B1410 (1964).

\bibitem{Aharonov2008} Y. Aharonov and L. Vaidman, ``The Two-State Vector Formalism: An Updated Review'', in \emph{Time in Quantum Mechanics}, Springer (2008).

\bibitem{Aharonov1988} Y. Aharonov, D. Z. Albert and L. Vaidman, ``How the Result of a Measurement Can Depend on the Future State of the System'', Phys. Rev. Lett. 60, 1351 (1988).

\bibitem{Cramer1986} J. G. Cramer, ``The Transactional Interpretation of Quantum Mechanics'', Rev. Mod. Phys. 58, 647 (1986).

\bibitem{Price1996} H. Price, \emph{Time's Arrow and Archimedes' Point}, Oxford University Press (1996).

\bibitem{Bell1964} J. S. Bell, ``On the Einstein-Podolsky-Rosen Paradox'', Physics 1, 195 (1964).

\bibitem{Freedman1972} S. J. Freedman and J. F. Clauser, ``Experimental Test of Local Hidden-Variable Theories'', Phys. Rev. Lett. 28, 938 (1972).

\bibitem{Aspect1982} A. Aspect, J. Dalibard and G. Roger, ``Experimental Test of Bell's Inequalities Using Time-Varying Analyzers'', Phys. Rev. Lett. 49, 1804 (1982).

\bibitem{Jackiw1972} R. Jackiw, ``Dynamical Symmetry of the Schrödinger Equation'', Phys. Today 25, 23 (1972).

\bibitem{Moshinsky1952} M. Moshinsky, ``Diffraction in Time'', Phys. Rev. 88, 625 (1952).

\bibitem{Garcia-Calderon1997} G. García-Calderón and A. Rubio, ``Diffusion, Diffraction and Tunneling in Time-Dependent Systems'', Phys. Rev. A 55, 3361 (1997).

\bibitem{Zurek1991} W. H. Zurek, ``Decoherence and the Transition from Quantum to Classical'', Phys. Today 44, 36 (1991).

\bibitem{Zurek2003} W. H. Zurek, ``Decoherence, Einselection and the Quantum Origins of the Classical'', Rev. Mod. Phys. 75, 715 (2003).

\bibitem{Bender2007} C. M. Bender, ``Making Sense of Non-Hermitian Hamiltonians'', Rep. Prog. Phys. 70, 947 (2007).

\bibitem{Mostafazadeh2002} A. Mostafazadeh, ``Pseudo-Hermiticity versus PT Symmetry'', J. Math. Phys. 43, 205 (2002).

\bibitem{Heller1995} E. J. Heller, ``Wavepacket Dynamics and Quantum Chaology'', J. Phys. Chem. 99, 2625 (1995).

\bibitem{Berry1977} M. V. Berry, ``Regular and Irregular Semiclassical Wavefunctions'', J. Phys. A 10, 2083 (1977).

\bibitem{Gutzwiller1990} M. Gutzwiller, \emph{Chaos in Classical and Quantum Mechanics}, Springer (1990).

\bibitem{Alicki2001} R. Alicki and M. Fannes, \emph{Quantum Dynamical Systems}, Oxford University Press (2001).

\bibitem{Robinett2004} R. W. Robinett, ``Quantum Wave Packet Revivals'', Phys. Rep. 392, 1 (2004).

\bibitem{Popescu1994} S. Popescu and L. Vaidman, ``Causality Constraints on Weak Measurements'', Phys. Rev. A 49, 4331 (1994).

\bibitem{Kwiat1991} P. Kwiat and R. Chiao, ``Observation of a Nonclassical Nature of Two-Photon Interference'', Phys. Rev. Lett. 66, 588 (1991).

\bibitem{Jennewein2001} T. Jennewein, G. Weihs, J.-W. Pan and A. Zeilinger, ``Experimental Nonlocality Proof'', Phys. Rev. Lett. 88, 017903 (2001).

\bibitem{Andersen2003} M. F. Andersen, A. Kaplan and N. Davidson, ``Echo Spectroscopy'', Phys. Rev. Lett. 90, 023001 (2003).

\bibitem{Kaplan2002} A. Kaplan, M. F. Andersen and N. Davidson, ``Suppression of Chaos in Quantum Kicked Systems'', Phys. Rev. A 66, 045401 (2002).

\bibitem{Valentini1991} A. Valentini, ``Signal-Locality, Uncertainty, and the Subquantum H-Theorem'', Phys. Lett. A 156, 5 (1991).

\bibitem{Leifer2017} M. S. Leifer and M. F. Pusey, ``Is a Time-Symmetric Interpretation of Quantum Theory Possible Without Retrocausality?'', Proc. R. Soc. A 473, 20160607 (2017).

\bibitem{Saunders2010} S. Saunders, J. Barrett, A. Kent and D. Wallace (eds.), \emph{Many Worlds? Everett, Quantum Theory, and Reality}, Oxford University Press (2010).

\bibitem{Dressel2014} J. Dressel, M. Malik, F. Miatto, A. N. Jordan and R. Boyd, ``Colloquium: Understanding Quantum Weak Values'', Rev. Mod. Phys. 86, 307 (2014).

\bibitem{Aharonov2016} Y. Aharonov, E. Cohen and E. Tollaksen, ``The Quantum Pigeonhole Principle'', PNAS 113, 532 (2016).

\bibitem{Kadanoff1966} L. P. Kadanoff, ``Scaling Laws for Ising Models near Tc'', Physics 2, 263 (1966).

\bibitem{Wilson1975} K. G. Wilson, ``Renormalization Group and Critical Phenomena'', Rev. Mod. Phys. 47, 773 (1975).

\bibitem{Cardy1996} J. Cardy, \emph{Scaling and Renormalization in Statistical Physics}, Cambridge University Press (1996).

\bibitem{Quach2017} J. Q. Quach, ``Which-way Double-slit Experiments and Born-Rule Violation'', Phys. Rev. A 95, 042129 (2017).
\end{thebibliography}
\end{document}